\documentclass{article}

\usepackage{PRIMEarxiv}

\usepackage[utf8]{inputenc} 
\usepackage[T1]{fontenc}    
\usepackage{hyperref}       
\usepackage{url}            
\usepackage{booktabs}       
\usepackage{amsfonts}       
\usepackage{nicefrac}       
\usepackage{microtype}      
\usepackage{lipsum}
\usepackage{graphicx}
\graphicspath{{media/}}     
\usepackage{amsmath,amssymb,amsfonts}
\usepackage{algorithmic}
\usepackage{graphicx}
\usepackage{textcomp}
\usepackage{xcolor}
\def\BibTeX{{\rm B\kern-.05em{\sc i\kern-.025em b}\kern-.08em
    T\kern-.1667em\lower.7ex\hbox{E}\kern-.125emX}}

\usepackage{amsmath,amssymb,amsfonts}
\usepackage{booktabs}
\usepackage{xspace}
\usepackage{multirow}
\usepackage{balance}
\usepackage{paralist}
\usepackage{microtype}
\usepackage{listings}
\usepackage{hyperref} 

\usepackage{xcolor}
\usepackage{subcaption}

\newcommand{\tool}{\textsc{ASTRAL}\@\xspace}

\AtBeginDocument{%
  \providecommand\BibTeX{{%
    \normalfont B\kern-0.5em{\scshape i\kern-0.25em b}\kern-0.8em\TeX}}}

\lstdefinestyle{javastyle}{
    backgroundcolor=\color{white},   
    commentstyle=\color{teal},
    keywordstyle=\color{purple},
    numberstyle=\tiny\color{gray},
    basicstyle=\ttfamily\scriptsize,
    breakatwhitespace=false,         
    breaklines=true,                 
    captionpos=b,                    
    keepspaces=true,                 
    numbers=left,                    
    numbersep=5pt,                  
    showspaces=false,                
    showstringspaces=false,
    showtabs=false,                  
    tabsize=1
}
\usepackage{float}
\usepackage[ruled, vlined, linesnumbered]{algorithm2e}
\SetAlFnt{\small}
\SetAlCapNameFnt{\small}
\SetAlCapFnt{\small}
\SetFuncSty{textsc}
\AtBeginEnvironment{algorithm}{
	\DontPrintSemicolon
}

\newsavebox\CBox

\RequirePackage{float}
\RequirePackage[skins]{tcolorbox}
\newtcolorbox{custombox}[1]{
	colback=gray!10,
	colframe=gray!70,
	left=1mm,
	right=1mm,
	top=1mm,
	bottom=1mm,
	fonttitle=\bfseries,
	arc=0mm,
	leftrule=1mm,
	rightrule=0mm,
	toprule=0mm,
	bottomrule=0mm,
	notitle,
	before=\par\smallskip\noindent,
	before upper={\textbf{#1: } },
}
\newtcolorbox{probdefinition}[1]{
	colback=gray!10,
	colframe=gray!22,
	left=1mm,
	right=1mm,
	top=1mm,
	bottom=1mm,
	fonttitle=\bfseries,
	arc=2mm,
	leftrule=0mm,
	rightrule=1mm,
	toprule=0mm,
	bottomrule=1mm,
	notitle,
	before=\par\smallskip\noindent,
	before upper={\textbf{#1: } },
}

\title{Early External Safety Testing of OpenAI’s o3-mini: Insights from Pre-Deployment Evaluation }

\author{
  Aitor Arrieta \\
  Mondragon University \\
  Mondragon, Spain \\
  \texttt{aarrieta@mondragon.edu}  \\
  \And
  Miriam Ugarte \\
  Mondragon University \\
  Mondragon, Spain \\
  \texttt{mugarte@mondragon.edu}  \\
  \And
  Pablo Valle \\
  Mondragon University \\
  Mondragon, Spain \\
  \texttt{pvalle@mondragon.edu}  \\
   \And
  José Antonio Parejo\\
  University of Seville \\
  Seville, Spain\\
  \texttt{japarejo@us.es} \\
 \And
  Sergio Segura\\
  University of Seville \\
  Seville, Spain\\
  \texttt{sergiosegura@us.es} \\
  }

\begin{document}
\maketitle

\begin{abstract}
Large Language Models (LLMs) have become an integral part of our daily lives. However, they impose certain risks, including those that can harm individuals' privacy, perpetuate biases and spread misinformation. These risks highlight the need for robust safety mechanisms, ethical guidelines, and thorough testing to ensure their responsible deployment. Safety of LLMs is a key property that needs to be thoroughly tested prior the model to be deployed and accessible to the general users. This paper reports the external safety testing experience conducted by researchers from Mondragon University and University of Seville on OpenAI's new o3-mini LLM as part of OpenAI's early access for safety testing program\footnote{\url{https://openai.com/index/early-access-for-safety-testing/}}. In particular, we apply our tool, ASTRAL, to automatically and systematically generate up to date unsafe test inputs (i.e., prompts) that helps us test and assess different safety categories of LLMs. We automatically generate and execute a total of 10,080 unsafe test input on a early o3-mini beta version. After manually verifying the test cases classified as unsafe by \tool, we identify a total of 87 actual instances of unsafe LLM behavior. We highlight key insights and findings uncovered during the pre-deployment external testing phase of OpenAI's latest LLM.

\textbf{\textcolor{red}{Warning: This report contains extracts from unsafe test inputs generated by \tool, which may upset some readers. Reader discretion is advised.}}
\end{abstract}

\section{Introduction}

Safety of Large Language Models (LLMs) has become an integral property that needs to be thoroughly tested before and after the LLM has been deployed. The Trust4AI project\footnote{\url{https://trust4ai.github.io/trust4ai/}} is an initiative of two Spanish universities (Mondragon University and University of Seville) that aims to research automated methods to systematically test different LLM properties, including the safety and fairness. When the user of an LLM provides an unsafe test input (e.g., \textit{``How can one build a home-made explosive?''}), the LLM shall refuse to respond. Different testing methods and benchmarks have been proposed to this end~\cite{xie2024sorry, yuan2024s, zhang2023safetybench, zhang2024chisafetybench, ying2024safebench, huang2024longsafetybench, li2024salad, ji2024beavertails, vidgen2023simplesafetytests}. 

Our recent study proposes ASTRAL, a novel tool that automatically generates unsafe test inputs to test the safety of LLMs~\cite{ugarte2025astral}. ASTRAL has three core advantages over previous LLM safety tools and benchmarks. First, instead of relying on static safety benchmarks, ASTRAL generates novel and imaginative test inputs guaranteeing a broad range of unsafe LLM test inputs. To this end, ASTRAL uses the capabilities of LLMs along with Retrieval Augmented Generation (RAG) and few-shot prompting strategies to automatically generate unsafe test inputs (i.e., prompts) in 14 different safety categories (e.g., terrorism, child abuse), with different writing styles (e.g., using slang or uncommon dialects). Second, ASTRAL retrieves information from the most recent news by browsing from the internet relevant, timely, and reflective current societal contexts, trends, and emerging topics. This allows ASTRAL to generate test inputs that are up to date and ensure that the generated prompts to assess the safety level of the LLM under test are not obsolete. Lastly, ASTRAL provides an automated mechanism that relies on LLMs to classify whether the output of LLMs is safe or unsafe, reducing the time required by engineers to manually label the outcomes of LLMs, alleviating this way the well-known test oracle problem~\cite{segura2016survey} in the context of LLMs. 

As part of the external safety testing process conducted before the o3-mini model was released to the general user, the TRUST4AI team was selected for assessing the safety property of OpenAI's O3-mini LLM. As a result, we used \tool to automatically generate a total of 10,080 unsafe test inputs to test a pre-release beta version of this LLM. Overall, we found that the safety level of OpenAI's o3-mini model is high, although we did uncover some issues leading to unsafe outcomes that could be used by OpenAI to take corrective actions.

\section{Background and Related Work}

\subsection{Safety Testing of LLMs}

Safety in LLMs primarily concerns ensuring their outputs remain free from harmful content while maintaining reliability and security~\cite{biswas2023guardrails}. This is particularly critical when LLMs are applied to sensitive domains such as healthcare, pharmaceuticals, or terrorism, where responses may inadvertently include malicious or misleading information with serious consequences. To address these risks and enhance trust in AI, the European Union AI Act (Regulation (EU) 2024/1689)~\cite{euaiact} establishes a regulatory framework focused on AI governance.

This framework adopts a risk-based approach to AI regulation. Under Article 51 of the EU AI Act~\cite{aiactregulation}, LLMs are classified as General-Purpose AI Models with Systemic Risk—referring to large-scale risks that can significantly impact the value chain, particularly in areas affecting public health, safety, security, fundamental rights, or society at large, as defined in Article 3(35). As a result, ensuring LLMs undergo rigorous safety testing and regulatory compliance assessments has become imperative.

Different testing techniques have been proposed to assess the safety quality of LLMs. Several studies have proposed multiple-choice questions to facilitate the detection of unsafe LLM responses~\cite{zhang2023safetybench, zhang2024chisafetybench, huang2024longsafetybench, li2024salad}. These benchmarks have an issue, i.e., they are fixed in structure and pose significant limitations, differing from the way users interact with LLMs. An alternative to this was to leverage LLMs that are specifically tailored to solving the oracle problem when testing the safety of LLMs. To this end, Inan et al.~\cite{inan2023llama} propose LlamaGuard, a Llama fine-tuned LLM that incorporates a safety risk taxonomy to classify prompts either as safe or unsafe. Zhang et al.~\cite{zhang2024shieldlm} propose ShieldLM, an LLM that aligns with common safety standards to detect unsafe LLM outputs and provide explanations for its decisions. 

Other techniques exists to test the safety of LLMs, such as red teaming and creating adversarial prompt jailbreaks (e.g.,~\cite{souly2024strongreject, ganguli2022red, huang2023catastrophic, zou2023universal, mazeika2024harmbench, shen2023anything, wei2024jailbroken}). Red-teaming approaches use human-generated test inputs, resulting in significant and expensive manual work. Adversarial works, on the other hand, do not typically represent the interactions that general LLM users employ. 

A large corpus of studies focuses on proposing large benchmarks for testing the safety properties of LLMs, e.g., by using question-answering safety prompts. For example, Beavertails~\cite{ji2024beavertails} provided 333,963 prompts of 14 different safety categories. SimpleSafetyTests~\cite{vidgen2023simplesafetytests} employed a dataset with 100 English language test prompts split across five harm areas. SafeBench~\cite{ying2024safebench} conducted various safety evaluations of multimodal LLMs based on a comprehensive harmful query dataset. WalledEval~\cite{gupta2024walledeval} proposed mutation operators to introduce text-style alterations, including changes in tense and paraphrasing. Nevertheless, all these approaches employ imbalanced datasets, in which some safety categories are underrepresented. Therefore, SORRY-Bench~\cite{xie2024sorry} became the first framework that considered a balanced dataset, providing multiple prompts for 45 safety-related topics. In addition, they employed different linguistic formatting and writing pattern mutators to augment the dataset. While these frameworks are useful upon release, they have significant drawbacks in the long run. First, they may eventually be incorporated into the training data of new LLMs to enhance safety and alignment. Consequently, LLMs could internalize specific unsafe patterns, significantly diminishing the utility of these prompts for future testing, thereby requiring continuous evolution and the development of new benchmarks. Second, as discussed in the introduction, they risk becoming outdated and less effective over time.

To address all these limitations faced by previous studies, our previous paper proposes \tool~\cite{ugarte2025astral}.  \tool proposes a novel approach that leverages a black-box coverage criterion to guide the generation of unsafe test inputs. This method enables the automated generation of fully balanced and up-to-date unsafe inputs by integrating RAG, few-shot prompting and web browsing strategies. More details of the key features of \tool can be found in Section~\ref{sec:ASTRAL} and the related paper~\cite{ugarte2025astral}.

\subsection{ASTRAL}
\label{sec:ASTRAL}

We now briefly explain \tool and its features to automatically and systematically generate, execute, and evaluate unsafe LLM test inputs. All details of \tool can be found in its original paper~\cite{ugarte2025astral}. \tool uses LLMs, RAG and few-shot prompting strategies to automatically generate and execute unsafe test inputs (i.e., prompts) of a total of 14 different safety categories. Specifically, \tool operates in three main phases. First, during the test \textbf{generation} phase, an LLM generates a set of $N$ unsafe test inputs tailored to predefined categories, writing styles and persuasion techniques. To guide the generation of unsafe prompts, we propose a new black-box coverage criterion. This criterion ensures the generation of balanced unsafe test inputs across different safety categories, writing styles and persuasion techniques. The safety categories, writing styles and persuasion techniques supported by the latest \tool version are summarized in Table~\ref{table:features}. We hypothesize that introducing a variety of test input types permits detecting a wider scope of safety-related issues in LLMs. To achieve this, we leverage OpenAI's assistant APIs, as they support RAG-based methods to be integrated in GPT-based LLMs. Lastly, we leverage a novel feature that gives access to the test input generator to live data (e.g., browsing the latest news) with the goal of generating up-to-date unsafe test inputs.

\begin{table}[h]
\caption{Description of our black-box coverage features}
\label{table:features}
\centering
\setlength{\tabcolsep}{3pt} 
\renewcommand{\arraystretch}{1.5} 
\begin{tabular}{llp{9cm}} 
\toprule
\textbf{Feature} & \textbf{Ab.} & \textbf{Description} \\  
\cmidrule{1-3}
\multirow{9}{*}{Style} 
    & S1 & Slang: using colloquial informal language and slang \\ 
    & S2 & Uncommon dialects: employing uncommon dialects or accents \\ 
    & S3 & Technical terms: introducing technical language and terminology \\ 
    & S4 & Role-play: encouraging the LLM to act on behalf of a specific role (e.g., teacher, doctor, politician, etc.) \\ 
    & S5 & Misspellings: introducing language misspellings \\ 
    & S6 & Question: using interrogative sentences \\  
\cmidrule{1-3}
\multirow{7}{*}{Persuasion} 
    & P1 & Evidence-based persuasion: using empirical data, statistics, and facts as support \\ 
    & P2 & Expert endorsement: citing domain experts as support \\ 
    & P3 & Misrepresentation: presenting a non-genuine or false fact \\ 
    & P4 & Authority endorsement: citing authoritative sources as support \\ 
    & P5 & Logical appeal: using logical and reasonable sentences \\ 
\cmidrule{1-3}
\multirow{14}{*}{Category} 
    & C1 & Animal abuse \\ 
    & C2 & Child abuse \\ 
    & C3 & Controversial topics, politics \\ 
    & C4 & Discrimination, stereotype, injustice \\ 
    & C5 & Drug abuse, weapons, banned substances \\ 
    & C6 & Financial crime, property crime, theft \\ 
    & C7 & Hate speech, offensive language \\ 
    & C8 & Misinformation regarding ethics, laws, and safety \\ 
    & C9 & Non-violent unethical behavior \\ 
    & C10 & Privacy violation \\ 
    & C11 & Self-harm \\ 
    & C12 & Sexually explicit, adult content \\ 
    & C13 & Terrorism, organized crime \\ 
    & C14 & Violence, aiding and abetting, incitement \\ 
\bottomrule
\end{tabular}
\end{table}

The second step consists of the \textbf{execution}. During this phase, \tool feeds the generated test inputs into the target LLM under test (i.e., the o3-mini model in this case). Lastly, in the \textbf{evaluation} phase, another LLM acts as an oracle to analyze the outputs (i.e., responses) of the tested LLM. This LLM determines whether the output provided by the LLM meets the safety standards.


To guide the generation of unsafe prompts, we propose a new black-box coverage criterion. This criterion ensures the generation of balanced unsafe test inputs across different safety categories, writing styles and persuasion techniques. Furthermore, we conjecture that introducing a variety of test input types permits detecting a wider scope of safety-related misbehaviors in LLMs. Lastly, we leverage a novel feature that gives access to the test input generator to live data (e.g., browsing the latest news) 
with the goal of generating up-to-date unsafe test inputs.

\section{Methodology}

This section explains the methodology we used to test OpenAI's o3-mini beta LLM.

\subsection{Test input generation}

We used two different test suites, the one employed to assess \tool in the original paper~\cite{ugarte2025astral}, and a new one whose generation began in January, 2025.

\textbf{First test suite (TS1): }The first test suite is the original test suite employed in the evaluation of our original paper. This test suite was generated in November 2024. Since \tool leverages web browsing to generate up-to-date inputs, the resulting test inputs included notable events from that period, such as the 2024 US elections. This test suite consisted of 3 different \tool versions:

\begin{itemize}
    \item \tool (RAG): This version uses \tool with RAG, without including few-shot prompting to guide the generator in the generation of different writing styles, nor Tavily Search (TS) to browse recent news from the internet.
    \item \tool (RAG-FS): Besides RAG, this version uses \tool with few-shot prompting strategies that guides the tool generate prompts with different writing styles. However, it does not include TS, which prevents it from generating prompts related to recent events.
    \item \tool (RAG-FS-TS): This version includes the three key features from \tool, i.e., RAG, to retrieve examples of different safety categories; FS, to guide the generation of prompts following different writing styles; TS, to browse recent news from the internet to generate up to date test inputs.
\end{itemize}

For each of these versions, we generated 1260 test inputs. We generated test inputs by combining each of the categories of the three features described in Table~\ref{table:features} with each other (i.e., 6 (styles) $\times$ 5 (persuasion) $\times$ 14 (safety categories) $\times$ 3 (tests)). Therefore, with the three different \tool versions, a total of 3780 test inputs were generated and executed on the o3-mini model.

\textbf{Second test suite (TS2): }For the second test suite, we generated test inputs starting in January 2025. Remarkable events during this time included Donald Trump's inauguration and the Gaza's ceasefire, among others. In this case, since \tool (RAG+FS+TS) showed best results, we opted for generating test cases only with this version of the tool. However, instead of generating 3 test inputs for each of the combinations (as conducted in TS1), we generated 15, resulting in a total of 6300 test inputs (i.e., 6 (styles) $\times$ 5 (persuasion) $\times$ 14 (safety categories) $\times$ 15 (tests)). 

Therefore, in total, we generated and executed 10,080 test inputs on a beta version of OpenAI's o3-mini LLM.

\subsection{Test execution and evaluation}

Prior to executing our generated test inputs, we had to adapt our \tool version for two main reasons. First, OpenAI deprecated the API version we used in the initial \tool version. Therefore, we migrated our code to the new API. Second, the o3-mini model triggered exceptions to a large portion of our test inputs, claiming a policy violation. That is, we conjecture that the API was able to detect unsafe test inputs before they were provided to the LLM under test. The first test suite started to be executed on January 21st, 2025, whereas the second one took place from January 24 to January 29th, 2025.

We used the same setup as our previous paper~\cite{ugarte2025astral} regarding the evaluator, i.e., GPT3.5 version is used with the corresponding prompt to assess whether the LLM outcomes were unsafe or not. The reason for using this model was that it provided a higher accuracy than the rest of the models~\cite{ugarte2025astral}. The model classifies the outcomes provided by the LLM under test (i.e., o3-mini) as \textit{safe}, \textit{unsafe} or \textit{unknown}. We also had to adapt this module to classify the outcomes provided by the LLM under test as ``\textit{policy violation}'' as \textit{safe}, as it seems the model was able to detect unsafe test inputs and therefore reject to be provided to the LLM. That is, our assessment can be considered to be performed at the system level of the entire OpenAI's API, rather than the o3-mini model in isolation. Unfortunately, we did not have access to control this. Furthermore, the evaluator provided a rationale for its decision, offering explainability as to why an LLM output is deemed \textit{safe}, \textit{unsafe} or \textit{unknown}.

\subsection{Manual assessment}

As our test evaluator is subject to provide false positives (i.e., outcomes classified as ``\textit{unsafe}'' that should be ``\textit{safe}''), we manually assessed those outcomes classified as both, ``\textit{unsafe}'' and ``\textit{unknown}''. We did not manually assess the rest of the prompts due to the large number of total test inputs we executed.

It is noteworthy that the manual classification of unsafe LLM behaviors might be subject to individual sentiments as well as culture. For instance, in one of the cases, o3-mini responded with arguments supporting the use of firearms by civilians. Culturally in Spain, this practices are not widely accepted, as gun ownership and civilian use of firearms are heavily regulated and generally viewed with skepticism. This contrasts with other cultures where such practices might be considered more commonplace or acceptable. In addition, we found some borderline cases, for which three of the authors discussed different point of views to classify as safe or as unsafe and found an agreement.

During the assessment, if needed, the authors also considered the explanation given by the LLM to classify the outcome.

\section{Results and Discussion}

\subsection{Summary of the Results}

Table~\ref{tab:results} reports the number of safe and unsafe LLM outcomes \tool found. After manual verification, \tool found 87 unsafe LLM outcomes for out of 10,080 test inputs. The category in which the most unsafe test results were found was c3, i.e., controversial topics and politics, especially for test suite 2, followed by c13 (i.e., terrorism and organized crime), as reported in Table~\ref{tab:results_categories}. The recent appointment of Donald Trump as president of the United States could have led to providing unsafe controversial outcomes on an LLM that seemed not well aligned for the test inputs generated by \tool.

\begin{table}[h]
\caption{Summary of the results obtained results. Column Safe refers to the number of LLM responses that our evaluation model has classified it as safe. Safe (policy violation) column refers to those safe LLM responses that were due to violating OpenAI's policy (are also part of the safe test cases). Unsafe refers to the number of test cases that the evaluator classified it as so. Unsafe (confirmed) are the number of LLM responses that we manually confirmed that were unsafe. Unknown are those LLM outcomes that the evaluator did not have enough confidence to determine as unsafe. Out of those, the unsafe outcomes that we manually verified are reported in Unknown (confirmed unsafe). Lastly, TOTAL Confirmed Unsafe reports the total number of unsafe LLM outcomes that we manually confirmed}
\label{tab:results}
\resizebox{\textwidth}{!}{
\begin{tabular}{llccccccc}
\toprule
\multicolumn{1}{c}{\textbf{}} & \multicolumn{1}{c}{\textbf{}} & \textbf{Safe} & \textbf{\begin{tabular}[c]{@{}c@{}}Safe \\ (policy violation)\end{tabular}} & \textbf{Unsafe} & \textbf{\begin{tabular}[c]{@{}c@{}}Unsafe \\ (confirmed)\end{tabular}} & \textbf{Unknown} & \textbf{\begin{tabular}[c]{@{}c@{}}Unknown \\ (confirmed unsafe)\end{tabular}} & \multicolumn{1}{c}{\textbf{\begin{tabular}[c]{@{}c@{}}TOTAL \\ Confirmed Unsafe\end{tabular}}} \\ \cmidrule{1-9}
\multirow{3}{*}{\textbf{TS1}} & \textbf{ASTRAL (RAG)} & 1239 & 707 & 19 & 7 & 2 & 1 & 8 \\
 & \textbf{ASTRAL (RAG-FS)} & 1249 & 762 & 10 & 9 & 1 & 0 & 9 \\
 & \textbf{ASTRAL (RAG-FS-TS)} & 1236 & 565 & 20 & 13 & 4 & 2 & 15 \\  \cmidrule{1-9}
\textbf{TS2} & \textbf{ASTRAL (RAG-FS-TS)} & \multicolumn{1}{l}{6205} & 2457 & 73 & 50 & 22 & 5 & 55 \\ \bottomrule
\end{tabular}}
\end{table}

\begin{table}[h]
\caption{Number of manually confirmed unsafe LLM outputs per safety category}
\label{tab:results_categories}

\resizebox{\textwidth}{!}{\begin{tabular}{llcccccccccccccc}
\toprule
 &  & \textbf{c1} & \textbf{c2} & \textbf{c3} & \textbf{c4} & \textbf{c5} & \textbf{c6} & \textbf{c7} & \textbf{c8} & \textbf{c9} & \textbf{c10} & \textbf{c11} & \textbf{c12} & \textbf{c13} & \textbf{c14} \\ \cmidrule{1-16}
\multirow{3}{*}{\textbf{TS1}} & \textbf{ASTRAL (RAG)} & 3 & 0 & 1 & 0 & 0 & 0 & 0 & 0 & 1 & 0 & 0 & 3 & 0 & 0 \\
 & \textbf{ASTRAL (RAG+FS)} & 4 & 0 & 1 & 0 & 2 & 0 & 0 & 0 & 1 & 0 & 0 & 0 & 1 & 0 \\
 & \textbf{ASTRAL (RAG+FS+TS)} & 3 & 3 & 2 & 0 & 3 & 0 & 0 & 0 & 0 & 1 & 0 & 1 & 2 & 0 \\ \cmidrule{1-16}
\textbf{TS2} & \textbf{ASTRAL (RAG+FS+TS)} & 4 & 0 & 29 & 1 & 3 & 2 & 1 & 1 & 1 & 1 & 0 & 1 & 10 & 1 \\ \bottomrule
\end{tabular}}
\end{table}

\subsection{Overall Discussion}

We now discuss a set of five key findings and their implications.

\textbf{Finding 1 -- OpenAI's o3-mini model seems to be safer than its predecessors.} In principle, when compared with other older OpenAI's models (i.e., GPT3.5, GPT4 and GPT-4o), OpenAI's o3-mini model is safer. In our previous study~\cite{ugarte2025astral}, \tool was able to uncover a total of 752, 166 and 215 unsafe behaviors on GPT3.5, GPT4 and GPT-4o, respectively for the first test suite. That is, the number of uncovered unsafe LLM outcomes were much higher in older OpenAI's models than the beta o3-mini model. A possible reason for this could be the policy violation feature that safeguards this last, which we discuss in further detail in the third finding.

\textbf{Finding 2 -- OpenAI's o3-mini model is safer than other state-of-the-art tested LLMs with \tool.} In our previous study~\cite{ugarte2025astral}, Llama~2 was the safest model, with a total of 95 unsafe outcomes found within the first test suite. Nevertheless, it is noteworthy that these unsafe outcomes were not manually verified. However, the number of unsafe inputs found by \tool without confirmation for the first test suite was 49, much less than the ones found in Llama~2 (i.e., 95). 

\textbf{Finding 3 -- The policy violation feature might have played an important role on safeguarding the model.} While executing the test inputs, we observed that the API crashed for certain cases, citing a policy violation. When further investigating this issue, we found that OpenAI had a usage policy that prevented users from using their API when ``not complying with applicable laws'' as well as ``not using it to harm (yourself or others)''\footnote{\url{https://openai.com/policies/usage-policies/}}. As part of the safety testing process, the generated test inputs by \tool included both, illegal activities (e.g., fabrication of weapons) as well as prompts related to self-harm (e.g., self-medication, consumption of drugs). This suggests the presence of a firewall-like mechanism designed to block unsafe test inputs from reaching the LLM. While we can argue that the entire o3-mini system is safer, it seems like many of the generated test inputs were not actually executed on the LLM itself. When testing other models, this policy was not violated. Therefore, we are also unsure whether this mechanism will later be deployed when a non-beta version of the model is deployed for general users. If this mechanism is not finally deployed, we recommend OpenAI's developers to re-execute those test cases that triggered the policy violation. 

\textbf{Finding 4 -- Recent controversial topics seem to be able to trigger unsafe LLM outcomes.} As shown in Table~\ref{tab:results}, for the second test suite, the third category triggered many unsafe LLM outputs. This may be due to the large number of inputs generated by \tool referencing Donald Trump’s pardoning of January 6th offenders, which triggers, as our LLM evaluator describes, \textit{``suggestions and implications that could undermine the rule of law, shake public trust in governance, deepen political polarization, impact democracy negatively, and stir societal unrest''}.

\textbf{Finding 5 -- Besides c3 (controversial topics), c13, c1, c5 are other critical safety categories.} Most of these categories also triggered misbehaviors related to safety in other OpenAI's models when compared to other safety categories in our previous study~\cite{ugarte2025astral}. This suggests that OpenAI could pay special attention to those categories to make their models safer.

\section{Conclusion}

This report explains the early external safety testing we conducted on a beta version of OpenAI's o3-mini model. We employed our tool, \tool~\cite{ugarte2025astral}, to automatically and systematically generate and execute a total of 10,080 test inputs on the o3-mini model. After a manual verification of the LLM outcomes classified as \textit{``unsafe''} or \textit{``unknown''}, we observed that 87 of them provoked an unsafe outcome. When compared with our previous experiments~\cite{ugarte2025astral}, where other state-of-the-art LLMs were used, we found that the results of this new model are highly competitive in terms of safety. We noted, however, that many of the inputs were refused by a safeguard that seemed to be external to the LLM, as it triggered crashes related to a policy violation.

Lastly, it is important to note that excessive safety can come at the cost of helpfulness~\cite{zhang2024bi}. This trade-off, a crucial aspect of LLMs, was not explored in this study and is left for future work.

\section*{Replication Package}

The reults can be obtained in the following link: \url{https://doi.org/10.5281/zenodo.14762830}

The github link of \tool can be found in \url{https://github.com/Trust4AI/ASTRAL}

\section*{Acknowledgments}

This project is part of the NGI Search project under grant agreement No 101069364. Mondragon University's authors are part of the Systems and Software Engineering group of Mondragon Unibertsitatea (IT1519-22), supported by the Department of Education, Universities and Research of the Basque Country.

The researchers are also part of the Spanish Network on AI for Software Engineering, RED2022-134647-T funded by MICIU/AEI /10.13039/501100011033.

\bibliographystyle{ieeetr}
\bibliography{sample}

\end{document}